\documentclass[12pt]{article}
   \usepackage{epsfig}
      \usepackage{amsmath}
      \usepackage{amssymb}
      \usepackage{caption}
\usepackage{subcaption}
\usepackage{cite}

\usepackage{epstopdf}

\usepackage{amsmath}

\usepackage{setspace}
  \usepackage{graphicx}
  \usepackage{color}

 \newcommand{\be}{\begin{equation}}
\newcommand{\bea}{\begin{eqnarray}}
\newcommand{\eea}{\end{eqnarray}}
\newcommand{\beq}{\begin{equation}}
 \newcommand{\ee}{\end{equation}}

\renewcommand{\>}{\rangle}
\newcommand{\<}{\langle}

\begin{document}
  \renewcommand{\theequation}{\thesection.\arabic{equation}}

\begin{titlepage}

  \bigskip\bigskip\bigskip\bigskip

  \bigskip

\centerline{\Large \bf {Bulk reconstruction and the Hartle-Hawking wavefunction}}

    \bigskip

  \begin{center}

 \bf { Daniel Louis Jafferis}
  \bigskip \rm
\bigskip

{\it  Center for the Fundamental Laws of Nature, Harvard University, Cambridge, MA, USA}
\smallskip

\vspace{1cm}
  \end{center}

  \bigskip\bigskip

 \bigskip\bigskip
  \begin{abstract}

In this work, a relation is found between state dependence of bulk observables in the gauge/gravity correspondence and nonperturbative diffeomorphism invariance. Certain bulk constraints, such as the black hole information paradox, appear to obstruct the existence of a linear map from bulk operators to exact CFT operators that is valid over the entire expected range of validity of the bulk effective theory.
By formulating the bulk gravitational physics in the Hartle-Hawking framework to address these nonperturbative IR questions, I will demonstrate, in the context of eternal AdS-Schwarzschild, that the problematic operators fail to satisfy the Hamiltonian constraints nonperturbatively. In this way, the map between bulk effective theory Hartle-Hawking wavefunctions and exact CFT states can be linear on the full Hilbert space.

 \medskip
  \noindent
  \end{abstract}

  \end{titlepage}

  \tableofcontents

\section{Introduction}

The gauge/gravity correspondence provides the best understood example of a complete theory of quantum gravity, that is, a precisely defined quantum system which is approximated by general relativity in the appropriate limit \cite{Maldacena:1997re, Gubser:1998bc, Witten:1998qj}. An aspect of this is that gravity is a non-renormalizable effective theory, whose ultraviolet completion is provided by string theory. However for the purposes of this note, it will be surprising infrared, albeit nonperturbative, constraints that will play the central role.

In recent  years, there has been considerable progress in elucidating the construction of bulk gravity observables in the dual CFT language. Some surprising phenomena have been found, particularly in situations involving causal horizons in the bulk. In particular, contradictions appear to arise between the existence of linear operators associated to bulk observables behind horizons and the semiclassical analysis around the expected smooth spacetimes.

A paradigmatic example is the information paradox of black hole evaporation \cite{Hawking:1976ra}. In the asymptotically AdS context, the evolution is obviously unitary by the duality with a unitary conformal field theory. Thus the paradox is recast as a problem in the existence of CFT operators that obey the expected bulk evolution equations and commutation relations in the weakly curved region behind the horizon that is relevant to the Hawking process.

These arguments were sharpened in the work of \cite{Mathur:2009hf,Almheiri:2012rt,Almheiri:2013hfa, Marolf:2013dba}, which demonstrated that exponentially small corrections in the Planck expansion cannot resolve the paradox. In particular, there cannot exist CFT operators that approximate the naively expected behavior of local bulk fields behind the horizon to within exponential accuracy in typical states of the black hole. It was suggested in \cite{Almheiri:2012rt} that this implies a sharp breakdown of the bulk effective theory even in certain regimes with low curvatures - that most black hole microstates have firewalls near the horizon.

The fascinating work of Papadodimas and Raju \cite{Papadodimas:2012aq, Papadodimas:2015xma,Papadodimas:2015jra}, and related work of \cite{Verlinde:2012cy, Verlinde:2013uja, Verlinde:2013vja, Verlinde:2013qya} and \cite{Guica:2014dfa}, showed how to explicitly construct bulk operators behind horizons in $1/N$ perturbation theory around a given approximately thermal pure state, using an analog of the KMS relation. The striking feature of their construction is that it depends on the background state, so that it is impossible to obtain a linear operator associated to the bulk observable that is valid on all microstates. This can then evade the AMPS(S) paradoxes. In the present work, I will focus for simplicity on the situation of a two sided black hole, where similar issues also arise \cite{Marolf:2012xe,Papadodimas:2015xma}.

Given that these interesting puzzles only appear when considering semiclassically distinct configurations, rather than in perturbation theory around a fixed background, it is appropriate to use a non-perturbative framework to describe the effective bulk physics. To that end, in this work, I will investigate the relation between the bulk Hartle-Hawking wavefunction \cite{Hartle:1983ai} and the dual QFT states. The asymptotically AdS arena is furthermore a promising one to make the Hartle-Hawking formalism more precise.

The main new feature that can be seen in such an analysis is that the observables whose existence as linear operators on the full Hilbert space would lead to inconsistencies are already ill-defined as operators on the wavefunction of the bulk effective theory. In particular, they are not non-perturbatively invariant under spacetime diffeomorphisms. I will show that they fail to obey certain non-perturbative consistency conditions analogous to the Wheeler-DeWitt equation \cite{DeWitt:1967yk} that implements the Hamiltonian constraints in perturbation theory.

One may define more refined, gauge fixed versions of these bulk operators that are non-perturbatively diffeomorphism invariant. These involve additional choices, and only agree with the naive observables on a small subspace of the full bulk effective theory Hilbert space. For these operators, it appears that no paradoxes arise, and they should be given by linear operators on the exact Hilbert space.

This paper is organized as follows. In section two, I sketch the basic logic of this work, and review the apparent obstruction to a linear map between certain bulk observables and CFT operators, in the context of eternal AdS-Schwarzschild. In section three, I argue using the Hartle-Hawking formalism that such observables fail to be nonperturbatively diffeomorphism invariant. In section four, I describe appropriately gauge fixed versions of bulk operators defined relationally to the boundary. Finally, in section five, I discuss possible implications for the information paradox and speculate about the appropriate framework for describing bulk measurements.

\section{Nonperturbative bulk reconstruction}

The relation between the Hilbert spaces of UV theories and their IR effective theories is, in general, one of projection. Thus one expects that the states of the IR Hilbert space can be mapped to states of the UV theory. They only span a subspace, since short distance modes remain in an adiabatic vacuum.

Consider the Wilsonian renormalization group flow in a quantum field theory on a $d$ dimensional spatial lattice, obtained by passing to an effective description with twice the lattice spacing. The Hilbert space of a $2^{\otimes d}$ block must be replaced by the Hilbert space of a single lattice site, of dimension $M$ \cite{White:1992zza, Kadanoff:1966wm}. This is done by a projection map, $$\otimes_{i_1, \dots i_d=1}^2 {\cal H}_{i_1, \dots i_d} \rightarrow {\cal H}_\textrm{block} ,$$ from the $M^{2^d}$ dimensional space to an $M$ dimensional space. The cokernel of this map should be interpreted as states in the microscopic theory in which no short distance modes are excited. Clearly information is lost during each step of the rg block transformation, and the projection must be chosen judiciously to avoid projecting out physically important states.

The Hilbert space of the long distance effective theory, ${\cal H}_\textrm{EFT}$, is the subspace of the full Hilbert space of the microscopic theory in which only long wavelength degrees of freedom are excited. This should be sharply contrasted with the subspace of states with total energy below some cutoff (in an IR finite system with a discrete spectrum). The latter would contain states with small numbers of ultra high energy quanta, which are not part of the effective theory, and would fail to contain coherent states of long wavelength modes with large total energy. For the present purpose of examining nonperturbative constraints in the bulk, it is important that ${\cal H}_\textrm{EFT}$ does contain such coherent states, the quantum analogs of low curvature solutions of the non-linear classical equations of motion.

In standard constructions, there is a fixed projection in the above definition of ${\cal H}_\textrm{EFT}$, which, for example, might select states of lowest energy density. An alternative possibility is the use of state dependent projections, in other words, a non-linear map. For example, in the MERA block transformation, the Hilbert space is projected onto the subspace with maximal entanglement with the rest of the system \cite{Vidal:2007hda}. This type of rg flow for states has proven to have significant computational utility in constructing ground states of certain many body systems. However the physical meaning of the state dependence of the resulting effective theory observables is unclear. In particular, if it were significant in some situations, it would seem to violate the linearity of quantum mechanics, which we certainly do not expect in long distance effective observables in condensed matter systems.

It is therefore important in the gravitational context to ask what a fundamental state dependence of bulk observables would imply for the Hilbert space of the effective bulk theory. In particular, if the theory in AdS is dual to a large $N$ conformal field theory, what is the relation between the exact CFT Hilbert space and the bulk effective theory Hilbert space?

It is essential to describe the bulk theory in a non-perturbative framework, since the constraints and paradoxes under discussion that motivate state dependent constructions only appear when comparing states that differ non-perturbatively. Perturbative treatments only refer to the behavior of observables in the subspace that can be obtained by acting with parametrically less than $N$ single trace operators on a given state with a known gravity dual. The CFT expressions for bulks fields around the AdS vacuum \cite{Banks:1998dd,Hamilton:2006az} and in other situations where they can be constructed using the bulk evolution equations fall into this category. Even if no causal reconstruction is possible, all constructions clearly lead to linear operators \cite{Almheiri:2014lwa} in such code subspaces. The question at hand is how these can be patched together over the full space of configurations described by the bulk effective theory.

State dependence of bulk observables in the emergent bulk gravity theory would seem to imply that there does not exist a linear map between the full bulk low energy effective Hilbert space and the exact Hilbert space. Such maps would then only exist for subspaces of the bulk Hilbert space defined perturbatively around a given state. A scenario of this type would be that a MERA construction of the CFT states determines the bulk state. Since the projection in MERA is done on maximally entangled subspaces, the result would be a non-linear map on the Hilbert space.

I will argue instead that the observables in question fail to be non-perturbatively diffeomorphism invariant. In other words, such observables do not exist as linear operators even in the bulk gravity effective  theory. Thus there is no contradiction with the existence of a linear map, ${\cal H}_\textrm{bulk} \hookrightarrow {\cal H}_\textrm{CFT}$, between all physical, gauge invariant states of the low energy bulk theory and states of the CFT. The existence of such a linear map is completely natural, given the AdS/CFT duality of the quantum systems. Furthermore, it appears that there is no need to restrict ${\cal H}_\textrm{bulk}$ to be smaller than the expected domain of validity of the bulk effective theory.

These situations must be examined in a non-perturbative long wavelength bulk effective description that appropriately imposes the constraints of temporal diffeomorphism invariance, such as the Wheeler-DeWitt wavefunction. That formalism can be made more precise in the asymptotically AdS context, as I will discuss in section 3. It is non-perturbative because it allows discussion of configurations that are significantly different than the vacuum (for example, states related to solutions of the classical non-linear equations). It is a low energy effective formalism in the bulk because short distance modes are not excited.

The Hartle-Hawking wavefunction is defined by a euclidean path integral up to a slice with on which the spatial metric and other bulks fields, collectively denoted by $h$, are fixed. This data should be understood as defined up to spatial diffeomorphisms. The temporal components of the metric at the slice, the lapse and the shift, are integrated over. This is because they appear in the action without any time derivatives. Given a state $|\psi\>$, this path integral defines the wavefunction, $\Psi(h)$, giving a linear functional on states of the bulk Hilbert space, $h: {\cal H}_\textrm{bulk} \rightarrow \mathbb{C}$.

An important feature of the AdS arena is that adjusting asymptotic sources in the euclidean path integral produces many states, as opposed to the situation in cosmological global de Sitter spacetime, where only the Hartle-Hawking no boundary state \cite{Hartle:1983ai} is naturally defined by a euclidean path integral. The states obtained by euclidean path integrals with sources for single trace operators generate the bulk long distance effective Hilbert space, ${\cal H}_\textrm{bulk}$.

By the standard relationship between the path integral and the Hilbert space in quantum mechanics, the inner product between two states $|\psi\>$ and $|\psi'\>$ is obtained by performing the euclidean gravity path integral with the asymptotic sources that defined the two states. Similarly, the overlap $\< h| h' \>$  between two of the Hartle-Hawking kets is given by the euclidean path integral with intrinsic metric $h$ and $h'$ on two spatial slice boundaries. In the asymptotically AdS context, the two noncompact slices must asymptotically join at the AdS boundary.

 A crucial feature is that because no gauge for temporal diffeomorphisms has been fixed, the overlaps between the kets, $M_{h h'} = \< h| h' \>$, are generally nonzero, and the kets do not form a basis: they are overcomplete. This simply reflects the fact that there are many spatial slices of the same spacetime geometry.

The ket states $|h \>$ are thus not linearly independent. For this reason, the matrix elements of any gauge invariant operator $A_{h h'} = \langle h | {\cal O} | h' \rangle$ must obey certain constraints. In the perturbative analysis, for $h' = h +\delta h$, Hartle and Hawking showed that these conditions imply that the $A_{h h'}$ satisfy the Wheeler-DeWitt equation, in other words, $A$ commutes with the Hamiltonian constraints.

The euclidean path integral additionally allows one in principle to determine the overlap between arbitrarily  different kets, which would be challenging to understand by integrating the Wheeler-DeWitt equation. As I will explain in more detail in section 3, the resulting conditions are nonperturbative analogs of the Hamiltonian constraints. These will be the key to the present analysis.

One way of understanding the state dependence required in the constructions of bulk operators is that naively distinct bulk states are not linearly independent. In other words, the code subspaces based on different background states have unexpected overlaps. For example, using the duality with the exact CFT, \cite{Papadodimas:2015xma,Papadodimas:2015jra} demonstrated that very long time evolution of the two sided AdS black hole results in states whose overlap does not decay to 0, but rather remains finite, albeit exponentially small in $N$.

The main point of this note is that this overlap of seemingly distinct states has the same origin as the lack of independence of the Hartle-Hawking kets. For this reason it is already a feature of the bulk effective theory Hilbert space, and the associated constraints on operators are imposing non-perturbative diffeomorphism invariance.

To illustrate the idea, consider the superselection sector of quantum gravity with two asymptotically AdS$_{d+1}$ boundaries. The spacetime geometry may be connected or not, depending on the state. However, the full quantum system consists of two decoupled copies of a large $N$ CFT. The CFTs are non interacting because the Hamiltonian density in gravity is a total derivative, hence the total Hamiltonian consists of two decoupled boundary terms, $H = H_L + H_R$, even when the spacetime is connected \cite{Maldacena:2001kr}.

A particularly straightforward example of a non-perturbative IR paradox appears in this context, as pointed out by Marolf and Wall \cite{Marolf:2012xe}. Consider a right framed bulk observable, such as a field operator at a position defined by a fixed proper distance, relative to empty AdS, along a geodesic that extends from a point on the right boundary at a given direction.

In any factorized state, this observable should be represented by a purely right CFT operator. After all in that case the left CFT is neither interacting nor entangled with the right one, so it seems that from the right point of view it cannot have any effect.

On the other hand, the thermofield double state is dual to the eternal AdS-Schwarzschild  geometry \cite{Maldacena:2001kr}, described by the metric $$ds^2 = - f(r) dt^2 + \frac{dr^2}{f(r)} + r^2 d\Omega_{d-1}^2, \textrm{where } f(r)=r^2+1-\frac{8 G_N M \Gamma(d/2)}{(d-1) \pi^{(d-2)/2)} r^{d-2} }.$$ The most salient feature of that spacetime  is the Einstein-Rosen bridge connecting the two sides. Thus the right framed bulk observable can enter the causal domain of the left boundary, and the observable cannot commute with all left operators in that state.

This contradicts linearity of the bulk observable, because the thermofield double state is a linear combination of factorized states, as we know from the microscopic CFT description, $$|\textrm{tfd} \> = \frac{1}{\sqrt{Z_\textrm{thrm}(\beta)}} \sum_E e^{-\beta E/2} |E\>_L \ |E\>_R.$$ In  \cite{Marolf:2012xe}, it is suggested that the AdS/CFT duality is incomplete in this situation and must be supplemented by a choice of connected or disconnected dictionary. In this work, I will argue instead that the bulk observables do not define non-perturbatively diffeomorphism invariant operators on the bulk effective Hilbert space. One must make additional choices in specifying the observable to make it gauge invariant, which determine whether it acts non-trivially on the left CFT. These play the role of the choice of dictionary, but we will see that they are already required to have a well-defined action on the Hartle-Hawking states in the bulk effective theory.

It is important to note that small corrections to the bulk gravity behavior of these operators cannot resolve this paradox. In particular, using arguments similar to those in \cite{Mathur:2009hf,Almheiri:2012rt, Almheiri:2013hfa}, the works \cite{Papadodimas:2015xma, Papadodimas:2015jra} demonstrated that any linear operator must have order 1 disagreement with the expected properties of a right framed bulk field behind the horizon on most states that are dual to time shifted versions of the eternal black hole.

The evolution of the thermofield double state by the left Hamiltonian is simply the application of a large diffeomorphism that acts nontrivially at  the left boundary. This makes it clear that such states, $|\psi_T\> = e^{i H_L T} |\textrm{tfd}\>$, should remain in the domain of validity of the bulk effective theory, even for long time scales $T = {\cal O}(e^S)$. This time shifted state is geometrically the same spacetime manifold as the eternal black hole, merely with a different origin of time on the left boundary. Thus any observables that are truly defined only relative to the right boundary should be unaffected, and the region near the horizon has low curvature for all $t$.

A very simple way to see that exponentially small corrections to the observables can't help is to consider the bulk observable, $C$, that counts the number of connected components of space. Suppose it has an expectation value of $1 + {\cal O}(e^{-N})$ on the states $|\psi_T\>$ in which the two boundaries are connected by an Einstein-Rosen bridge, and an expectation value of $2 + {\cal O}(e^{-N})$ on factorized states, in which the spacetime consists of disconnected left and right pieces. Using the microscopic CFT description of $|\psi_T\>$ as $\frac{1}{\sqrt{Z_\beta}} \sum e^{(-\frac{\beta}{2} + i T) E} |E\>_L |E\>_R,$ one obtains a contradiction, because averaging the result for $|\psi_T\>$ over a sufficiently long interval of the parameter $T$ approximately projects onto individual energy eigenstates.

In more detail, if $\<E,E| C |E', E'\> = 2 \delta_{E-E'} + \epsilon(E, E')$, where $\epsilon$ is exponentially small in the black hole entropy, then $$\<\psi_T| C | \psi_T\>  = 2 + \frac{1}{Z_\textrm{thrm}(\beta)} \sum_{E, E'} e^{-\frac{\beta}{2}(E + E') + i T (E'-E)} \epsilon(E, E'). $$ It is impossible for the phases in the second term to add coherently for all $T$ over a interval of order $e^S$ in such a way that the double sum over the exponentially large number of states of the exponentially small quantities $\epsilon(E,E')/Z_\textrm{thrm}(\beta)$ is of order 1. Therefore,
for most $T$, the result will be the average value of 2 up to exponentially small corrections, in contradiction with its expected value of $1 + {\cal O}(e^{-S})$.

A sketch of the argument in \cite{Papadodimas:2015xma, Papadodimas:2015jra}  for the more interesting operators that measure a field at bulk points defined relationally to the right boundary is as follows. Consider such an observable, for example the particle number operator, $N$, constructed out of the right framed field operator, which has an ${\cal O}(1/S)$ expectation value in the thermofield double state. It  will then also have an ${\cal O}(1/S)$ expectation value in the time shifted states. Then one can average over exponentially long times to obtain $$\frac{1}{2T} \int_{-T}^T \< \psi_t| N |\psi_t\> dt = \sum_E \frac{e^{-\beta E}}{Z(\beta)} \<E,E| N |E,E\> + \sum_{E\neq E'} \frac{e^{-\beta (E+E')/2}}{Z(\beta)} \frac{\sin((E'-E) T)}{(E'-E)T} \<E,E|N|E',E'\>.$$

The second term becomes small when $T$ is large enough, which proves that the first term is also of order $1/S$. This is true for all temperatures $\beta$. Performing a Legendre transform, one can approximately project on to factorized states. It is then easy to show that this implies that the right framed operator must have a small expectation value in all factorized states of the same energy. This is inconsistent with the bulk predictions.

In the next sections, I will analyze this situation in the Hartle-Hawking formalism. It is applicable here because all of the time shifted states are well-described by wavefunctionals in the low curvature regime. In particular, one can find nice slices even in these exponentially time shifted states which have small intrinsic and extrinsic curvatures, if the black hole is large in Planck units.

Before turning to the more detailed analysis of these observables in the bulk Hartle-Hawking framework, it is interesting, although simple, to see that the nonzero overlap between a factorized like such as the vacuum $|0\>_L \ |0\>_R$ and the thermofield double state is already calculable in the gravity path integral. It is precisely the lack of orthogonality between these states with connected versus disconnected semiclassical descriptions which led to the paradox.

The euclidean gravity saddle that contributes to $$\<0,0| \textrm{tfd} \> = \frac{Z_{S^d}}{\sqrt{Z_{S^d}^2} \sqrt{Z_{S^{d-1} \times S^1}}} = \frac{1}{\sqrt{Z_\textrm{therm}}},$$ is just euclidean AdS$_{d+1}$. Note that in the CFT language, the vacuum caps that produce the state $|0,0 \>$ and the cylinder that produces $|\textrm{tfd}\>$ are conformally flat, so the overlap partition function is not only topologically but in fact conformally equivalent to $S^d$. The denominator is just the usual normalization of factor for the states produced by these path integrals.

\begin{figure}[h!]
\begin{center}
\vspace{5mm}
\includegraphics[scale=.25]{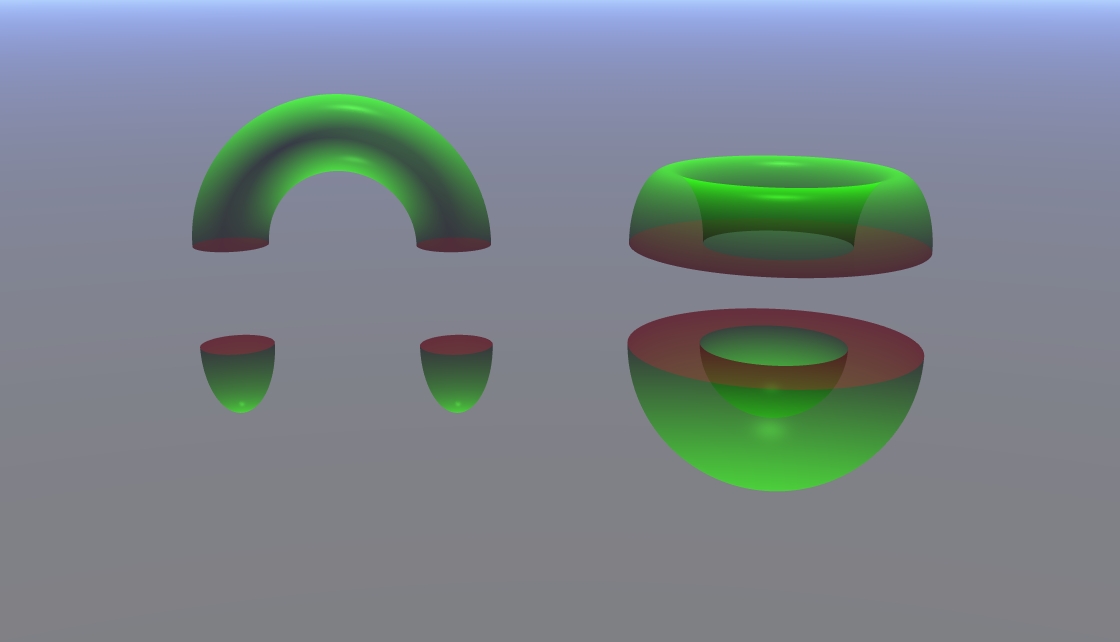}
\vspace{5mm}
\caption{The factorized vacuum state is produced by a pair of hemisphere asymptotic boundaries, while the black hole state is produced by an annulus asymptotic boundary. The metric is not the induced one associated to the figure. The left and right figures show the two gluings described in the text, with a disconnected versus a connected slice.}
\label{Topologyoverlap}
\end{center}
\end{figure}

The inner product between the vacuum and the thermofield double state in the gravitational path integral is shown in on the left in Figure 1. It can be interpreted as an amplitude that the thermofield double state is a disconnected geometry. That this is non-vanishing is not surprising, given that the disconnected configuration can even be dominant over the black hole (below the Hawking-Page phase transition). On the other hand, one can equally well glue together the bulk configurations shown on the right in Figure 1. Here it appears to be an amplitude that the vacuum is a connected spacetime.

A important aspect is that it would be completely wrong to add together these ``two'' possibilities. They are just different depictions of the same saddle, as can be seen in Figure 2, which is an alternative, but topologically equivalent, depiction of the two slicings of Figure 1. This is precisely because $C$, the number of connected components of space, is not a nonperturbatively diffeomorphism invariant operator. It depends on the choice of slice, and so fails to obey the Hamiltonian constraints non-perturbatively.

 As explained above, a paradox very similar to that described in \cite{Marolf:2012xe} appeared  for the putative observable $C$. By the AdS/CFT duality, it should be 2 on factorized states, and 1 on the thermally entangled states, which contradicts linearity. What we see now is that $C$ is simply not gauge invariant, so this indicates no obstruction to a linear map between ${\cal H}_\textrm{bulk}$ and ${\cal H}_\textrm{CFT}$.

\begin{figure}[h!]
\begin{center}
\vspace{5mm}
\includegraphics[scale=.35]{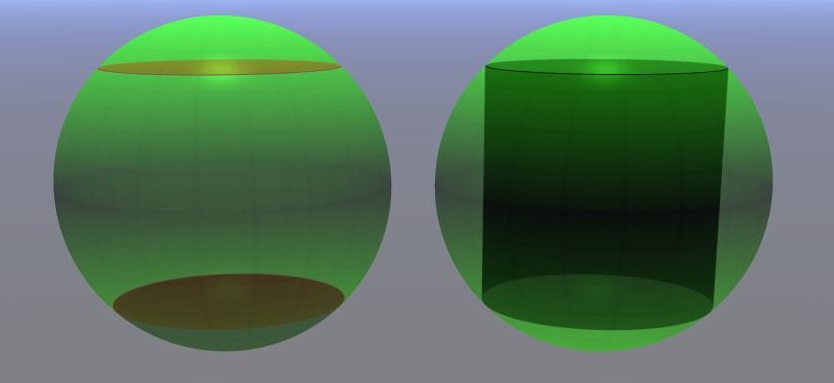}
\vspace{5mm}
\caption{Here it is clear that these are simply two different bulk slices of the same configuration, described by the euclidean AdS saddle. The left shows the disconnected slice and the right shows the connected slice. The slices are at $t=0$. The black hole state is produced by the annulus asymptotic boundary in the middle of the spheres, while the vacuum is produced by the cap boundaries at the top and bottom of the spheres.}
\label{Topologyoverlap2}
\end{center}
\end{figure}

The Hartle-Hawking kets with different topologies are not independent, and no complete linear operator $C$ can act on them with different eigenvalues. The nonzero overlap between these kets implies topology ambiguity, not dynamical change of the topology. A state that ``definitely'' has a given topology still has a nonzero probability to have a different one. This is particularly clear in the AdS context, since the global Hamiltonian makes no appearance in this discussion. If a related non-perturbatively gauge fixed operator was defined, then its dynamical evolution could be discussed.

  This fact is an important point of consistency in the proposal that entanglement encodes spatial connection \cite{VanRaamsdonk:2009ar, VanRaamsdonk:2010pw} by nontraversable wormholes \cite{Maldacena:2013xja}, since entanglement is not measurable by any operator. The nonexistence of a linear operator that measures the spatial topology has been pointed out by \cite{Bao:2015nca, Berenstein:2016pcx, Berenstein:2017abm} and in related work of \cite{Almheiri:2016blp}, based on the duality with the CFT. This is true even without the presence of horizons, as shown the examples of \cite{Berenstein:2016pcx, Berenstein:2017abm}. Now we see that this is already a feature of the bulk effective theory.

 It is more subtle to see that the geodesic defined operators are not gauge invariant beyond perturbation theory. As explained in section 4, it requires extra data to define them, so that the geodesic dressing lies in a space-like slice. This is necessary in order to have a straightforward action on the bulk Hilbert space. The end result is that appropriately diffeomorphism invariant versions of these observables do not lead to paradoxes.

\section{The Hartle-Hawking wavefunction in AdS}

Consider the collection of states that can be produced by performing a CFT path integral with sources for single trace operators. This class encompasses all states that are accessible in the bulk effective theory. Given any such state, we want to find its description as a wavefunctional $\Psi(h)$. Here $h$ represents the data of the bulk metric and fields on a spatial slice. They are required to obey  AdS asymptotic conditions that are described below, and we should consider $h$ to label equivalence classes with respect to the redundancies generated by spatial gauge transformations obeying the appropriate boundary fall off conditions.

The Hartle-Hawking prescription \cite{Hartle:1983ai} is to perform the euclidean path integral with sources on the AdS boundary, up to a slice with data $h$,
 \begin{equation} \Psi(h) = \int_{g|_{\partial M} = h,\ g|_\infty = {\cal J}} Dg \ e^{-S(g)}, \end{equation} up to a normalization constant, where the time cut boundary of space is $\partial M$, and the AdS asymptotics, $g|_\infty$, are schematically given by the sources ${\cal J}$ that define the state. In other words, one only integrates over metrics and bulk fields that obey the appropriate Fefferman-Graham fall-off conditions. Writing the metric near the slice as \begin{equation} ds^2 = (N^2-N_i N^i) dt^2 + 2 N_i dx^i dt + h_{ij} dx^i dx^j, \end{equation} where $x_i$ are coordinates on the slice, one integrates over the lapse, $N$, and shift, $N_i$, at the slice, keeping only $h_{ij}$ fixed.  One similarly integrates over all over bulk fields, with AdS asymptotics given by sources and fixed on the slice as $\varphi|_{\partial M} = \varphi(x)$ to obtain the wavefunctional $\Psi(h,\varphi)$.

 This formalism is general, but for concreteness, one may consider gravity described by the Einstein-Hilbert action, $S = \int_M \sqrt{g} (R-2 \Lambda) + 2 \int_{\partial M} \sqrt{h} K$, where in AdS $\Lambda <0$, and trace of the extrinsic curvature $K_{ij} = \frac{1}{N} \left(-\frac{1}{2} \frac{\partial h_{ij}}{\partial t} + D_{(i} N_{j)}  \right)$ appears in the boundary term. More precisely, the AdS asymptotic boundary should be cut off, and the action obtained as a limit after subtraction of the appropriate counter terms, as discussed in this context of producing Lorentzian states from euclidean sources on a cap in \cite{Skenderis:2008dh, Skenderis:2008dg}

 In AdS, the spacetime metric is required to have the asymptotic form $$ds^2 = \frac{L^2}{z^2} dz^2 + \frac{\eta^{\mu\nu} dy^\mu dy^\nu}{z^2} + \dots,$$ where $y^\mu$ are the boundary spacetime coordinates, $L$ is the AdS radius, and the subleading terms vanish in $z \rightarrow 0$ limit.

 This implies that the slice metric $h_{i j} d x^i d x^j = \frac{L^2}{z^2} dz^2 + \dots $. Moreover, the $z$ shift must obey $N_z \rightarrow 0 $ at the boundary.

 Only those spatial diffeomorphisms that vanish as $z \rightarrow 0$ are redundancies, and $\Psi$ is a functional of such equivalence classes of $h$. The asymptotic vanishing of $N_z$ implies that the ADM Hamiltonian is a nontrivial boundary operator, rather than a constraint.

 The path integral above should be understood as being evaluated in bulk perturbation theory around its saddles. It is still non-perturbative in that one may consider states and $h$ that differ classically, in other words, at leading order. One should only allow the metrics, $h$, on the slice which have no Planck scale features. More precisely, one must perform the path integral in a bulk effective theory with cutoff $\Lambda \ll M_\textrm{Pl}$, and the slice metric must live in the cutoff configuration space, with no wavelengths smaller than $\Lambda^{-1}$. If there is no bulk saddle with small curvatures that contributes, the result will be close to 0. Some CFT states may have small amplitude on all $h$, since they contain Planckian or string scale objects in the bulk. As discussed above, contributions from subleading euclidean saddles will lead to exponentially small corrections to the Hartle-Hawking kets, which will not significantly affect the conditions of non-perturbative diffeomorphism invariance.

 The bulk effective theory Hilbert space is spanned by states that can be produced by performing such euclidean path integrals with sources, as can be seen by analytic continuation of Lorentzian bulk configurations (in general the sources will be complex). The Hartle-Hawking path integral then defines a map from $H_{\textrm{bulk}} \longrightarrow \mathbb{C}$. Using the Hilbert space inner product defined on the states $|\psi\> \in {\cal H}_\textrm{bulk}$, one identifies $|h\>$ with a state (strictly speaking, a limit of states since these may have divergent normalizations). As usual, the relation between the exact Hilbert space and a low energy effective Hilbert space is by projection, $H_{CFT} \rightarrow H_{bulk}$. Thus we can also think of $|h\>$ as living in $H_{\textrm{CFT}}$, by applying the adjoint of the projection map.

 This set of kets is overcomplete, and the matrix of their inner products, $M(h,h') = \< h| h'\>$, given by the euclidean path integral between slices with intrinsic data $h$ and $h'$, are generally nonzero. For this reason is not entirely trivial to determine the inner product on the wavefunctionals $\Psi(h) = \<h|\psi \>$, in other words to find a $K_{h h'}$ such that $\< \psi_1 | \psi_2 \> = \int dh \ dh' \Psi_1(h)^* K_{h h'} \Psi_2(h')$. The kernel $K$ would simply be the inverse of $M$ if the set was linear independent. Instead, there are many matrices $K$ that satisfy $M K M = M$.

 One such kernel is a differential operator in the space of $h$, representing the inner product as $$\< \psi_1 | \psi_2 \> = \int \prod_{x} dh^{i j}(x) \frac{h_{i k} h_{j l} +h_{i l} h_{j k} - h_{ij} h_{kl}}{\sqrt{h}} \left[  \Psi_1(h)^* \left(\frac{\overleftarrow{\delta}}{\delta h_{k l}} - \frac{\overrightarrow{\delta}}{\delta h_{k l}}\right) \Psi_2(h) \right],$$ where the product is over all points in the spatial slice \cite{DeWitt:1967yk}. This results in an integration over spatial metrics that excludes the local Weyl factor, which is a timelike direction in the superspace of metrics. In this work, I will not need to use the inner product on the wavefunctionals.

An important caveat to this path integral based approach to the bulk effective theory is that the full rules for summing over different spacetime topologies in the gravity partition function are not known. For example, in a calculation in AdS$_{d+1}$, with asymptotic boundary $S^d$, one might including a sum over topologies with extra handles. Cutting these with a time slice could include a disconnected compact spatial region.

One of the puzzles is that  in the system with two asymptotic boundaries, the CFT implies that correlation functions in the vacuum, $\<0,0| {\cal O}_L {\cal O}_R |0,0\>$, factorize between the left and right. However, if connected spacetime saddles contributed, this factorization would appear mysterious on the gravity side. This issue arose in \cite{Yin:2007at} in the attempt to define the partition function of pure gravity in three dimensions and find its CFT dual. Note that in the example of the vacuum correlators for a 2d CFT on a pair of circles, there is no such saddle, since no smooth hyperbolic 3-manifold may have two disconnected $S^2$ boundaries. However, as discussed in \cite{Yin:2007at}, such euclidean saddles do exist with disconnected higher genus boundaries. In that work, it was proposed that the gravity path integral simply does not include them.

In spite of this gap in understanding, the discussion of the present paper is unaffected. The reason is that the issue at hand is about the different topologies of slices that cut a given spacetime saddle, and their ket overlaps. Thus the question of whether other spacetime saddles should be summed over is not important for the characterization of non-perturbative diffeomorphism constraints.

The Wheeler-DeWitt equation $$\left(- \frac{1}{2} h^{-1/2} (h_{i k} h_{j l} + h_{i l} h_{j k} -h_{i j} h_{kl} \frac{\delta^2}{\delta h_{ij} \delta h_{kl}} - R(h) h^{1/2} + 2 \Lambda h^{1/2} \right) \Psi(h) = 0,$$ where $R$ is the intrinsic curvature of the slice can be derived \cite{Hartle:1983ai} from the lack of independence of the kets $|h\>$, which implies that $\Psi(h) = \<h | \psi \>$ is a redundant data. The differential equation results from expansion of the kets under small variation, $|h + \delta h \>$.

There are also non-perturbative constraints that would be hard to see by integrating the Wheeler-DeWitt equation. In particular, there is a conceptually identical lack of independence between the data with different topologies of the spatial metric, $h$. The Hartle-Hawking procedure defines wavefunctions $\Psi(h_1)$, $\Psi(h_1,h_2)$, ... which each possible spatial topology, including the number of connected components. However the Hilbert space does not consist of separate factors, since the kets are not linearly independent.

Unlike in gauge theory, in gravity it is impossible to fix the gauge even in perturbation theory by imposing a condition on the spatial $h$. This is because the Wheeler-DeWitt equation is second order in $\delta/\delta h$.  Furthermore, standard gauges do not actually fix the gauge nonperturbatively.

In particular, the diffeomorphism group for different topologies is different. In the two sided system, the difference in the group of diffeomorphisms is associated to the existence of the gravitational analog of the Wilson line operators discussed in \cite{Harlow:2015lma, Guica:2015zpf}.

Consider the example of spacetimes with a pair of asymptotically AdS boundaries. The AdS/CFT duality implies that the kets with disconnected topologies $|h_1, h_2\>$ in fact span the entire Hilbert space. This is hard to see explicitly in the bulk effective theory analysis, since obtaining the formula for connected kets in terms of disconnected kets involves partially diagonalizing the matrix of overlaps, $M_{h h'}=\langle h| h' \rangle$. The result will depend on the details of the cutoff scheme $\Lambda$.

For a different class of supersymmetric horizonless geometries \cite{Lin:2004nb} in global AdS$_5$, it is possible to see the linear relations explicitly, using the explicit description of the dual CFT microstates \cite{Berenstein:2016pcx, Berenstein:2017abm}. Without this exact dictionary protected by supersymmetry, the bulk effective theory analysis would in general lead to a linear relation that was cutoff dependent. Nevertheless, the matrix, $M_{h h'}$, of kets overlaps is a UV safe quantity.

The states $|\textrm{top}, h\>$ are overcomplete (here $h$ represents a metric on the topological space labelled by the first argument), so it is mathematically inconsistent to define a linear operator by independently specifying its action on each of these states. But given a linear operator ${\cal O}$, one can compute its matrix elements $A(h_1, h_2) = \<\textrm{top1}, h_1 | {\cal O} | \textrm{top2}, h_2\>$. Then these must obey some relations. By considering small changes in $h$ with a fixed topology, one can show that $A$ must commute with the Hamiltonian constraints. Similarly, by considering different topologies, one can show that $A(h_1, h_2)$ obeys further conditions. These are nonperturbative constraint equations. A collection of such matrix elements that don't satisfy these constraints do not give a mathematically consistent linear operator; it is not gauge invariant under temporal diffeomorphisms.

The AdS/CFT duality implies that disconnected topology states span the Hilbert space of the two sided AdS system. Therefore defining the action of an operator ${\cal O}$ on those states {\it determines} its action on kets with other topologies. It will be perfectly well defined on the thermofield double state.

What is disallowed is to say that one defines an operator like $C$, the number of connected components of space, which obeys $C|h_1,h_2\> = 2 |h_1, h_2\>,\ C|h_1\> = 1 |h_1\>$. This is just as bad as writing a set of matrix elements that fail to obey the Hamiltonian constraints. Therefore there is no candidate operator in the bulk effective theory that counts the number of disconnected components of the spacetime.

The euclidean gravity path integral implies that naive basis states associated to different spatial topologies are not linearly independent - they are related by the constraint equations. Therefore the number of spatial components is not non-perturbatively diffeomorphism invariant. There is no associated operator on the full bulk effective Hilbert space that satisfies the non-perturbative constraint equations.

It is interesting to contrast this fact with the structure of the classical phase space of general relativity. The latter has distinct components with different spatial topology, since there are no non-singular topology changing solutions. Thus naive quantization of this phase space would seem to imply that they give rise to orthogonal subspaces of the Hilbert space. Moreover, the topology of space at a given asymptotic boundary time appears to be a diffeomorphism invariant property of nonsingular spacetimes.

Similarly, there is no straightforward Lorentzian interpretation of the euclidean instanton that describes the overlap of the factorized vacuum with the eternal two sided black hole. One perspective here is that semiclassical quantization and the path integral formulations of the bulk effective gravity theory disagree, and comparison with the exact CFT result is what that tells us that the path integral method is the correct procedure in this situation. Of course, the disagreement is non-perturbative in the $G_N$ expansion.

However, it is more precise to say that Lorentzian semiclassical quantization is agnostic about these issues, since the relevant solutions always involve singularities, which are outside the domain of validity of the bulk effective theory. Analogously, the infinitesimal version of the Hartle-Hawking ket overlaps $\langle h | h + \delta h \rangle$ are described by the Hamiltonian constraint equations with a straightforward Lorentzian interpretation, but it not easy to integrate up to see the overlap between kets with different topologies. This is because the slice has to become singular somewhere along any deformation which changes its topology, and again one exits the regime of validity of the long distance theory. The fact that using the euclidean path integral one may directly define and calculate the overlap of non-perturbatively distinct kets is a main virtue of the Hartle-Hawking formalism.

\section{Boundary relational operators}

In classical gravity in asymptotically AdS spacetimes, one may define relational observables, for example the value of a bulk field at some regularized distance, $z$, along a geodesic that extends transversely into the bulk from a fixed boundary point, $x$. These are fully gauge invariant, since AdS boundary conditions require that diffeomorphisms vanish at the boundary.

In perturbation theory around empty AdS in Poincare patch, one may think of such an observable $\varphi(x,z)$ as the bulk field in Fefferman-Graham coordinates. In general, this is not a good description since Fefferman-Graham is not a good gauge around general backgrounds. The geodesics from different boundary points can intersect, causing this coordinate system to break down.

In the quantum path integral, one can insert such an observable, since it is well-defined for any off-shell metric configuration. However, it is a novel feature of gravity that this does not automatically imply the existence of a well-defined operator on the physical Hilbert space. In quantum field theory, one can perform the path integral up to some time, act with an operator, and then continue the path integral, so there is a direct relation between insertions and operators. In gravity, there is no canonical choice of time slice, so this procedure does not work. For example, insertion of $\varphi(x,z)$ into the path integral defining the wavefunctional $\Psi(h)$ is not well-defined, since for some metrics, the geodesic will exit the slice $h$. One needs to appropriately gauge fix the bulk operators nonperturbatively, so that they are compatible with the slices.

If it was possible to pick a non-perturbatively good gauge for diffeomorphisms, then there would be a simple relationship between insertions in the path integral in that gauge and diffeomorphism invariant operators. One of the essential points is that no such gauge exists, as can be seen by the fact that the causal relations between observables depend on the spacetime.

Another way of describing the same feature is that there are multiple ways of preparing the same state from a euclidean path integral. For example, the thermofield double state can be produced by a path integral with cylinder boundary or via the microscopic relation $|\textrm{tfd} \rangle = \frac{1}{\sqrt{Z(\beta)}} \sum e^{-\beta E/2} |E\rangle_L \ |E\rangle_R$. In general, an insertion of a geodesically defined observable into a path integral involving one or the other of these need not give the same result. Therefore there is no well-defined action on states.

It is possible to define closely related operators that are non-perturbatively diffeomorphism invariant. But additional choices are required, and the resulting operators do not have the expected action of all states. The following is one such prescription.

Consider kets $|h, \textrm{GD}\>$ defined by doing the path integral only over metrics that have the Fefferman-Graham form along a single geodesic originating at the boundary at a point $x_*$ and terminating at a point in the interior at a fixed regulated distance, $z$, defined with respect to empty AdS. In other words, one restricts the integration over the radial component of the shift to vanish along this geodesic between the AdS boundary and the point defined by $z$, in addition to fixing the spatial gauge on the slice to have the Fefferman-Graham form along that geodesic. The path integral is thus  only over $d+1$ dimensional metrics whose restriction to the $t=0$ slice is given by
$$ds^2_{d+1} = h_{i j} dx^i dx^j + N_i d t dx^i + S dt^2,\textrm{ where }h_{z z}(x_*, z) = \frac{L}{z^2}, h_{z a} = 0, N_z=0\textrm{ for }0<z<z_*,$$ where the index $i$ runs over all spatial directions, $z$ and the boundary spatial directions, $a$.
This gives a well-defined linear functional on the Hilbert space, $h_\textrm{GD}: {\cal H}_\textrm{bulk} \rightarrow \mathbb{C}$. Note that there is no integration over metrics in which this geodesic is chronal (ie. no two points on this space-like geodesic may time-like related), so these kets will span a smaller subspace than the Hartle-Hawking kets.

 These kets are still not all linearly independent, but ones with different values of the fields along the special geodesic will be orthogonal. This is because there is no integration over the shift along the geodesic, thus in the path integral that computes $\<h_1, \textrm{GD}| h_2, \textrm{GD} \>$ will have the marked geodesic in the two slices identified. Therefore one can define a bulk field operator along the geodesic using these kets as
 $$\Phi_\textrm{GD}(x_*,z_*) |h, \textrm{GD}\> = \varphi(x_*,z_*) |h, \textrm{GD}\>.$$ This is well-defined because the definition of the restricted kets implies that they are orthogonal if the values of $\varphi$ at the point $(t=0, x_*, z_*)$ are different.

 In this way, one can define a nonperturbatively gauge fixed version of the bulk field. The action of this operator on other states is determined by the above definition, supplemented by the condition that it annihilates the orthogonal complement of the full Hilbert space.

 Note that in the eternal black hole, this particular observable never enters the region behind the horizon (in the classical approximation), since the geodesic is exactly radial and, if long enough, would pass through the bifurcation surface directly into the left causal wedge. That is sufficient to phrase the paradox \cite{Marolf:2012xe} reviewed in section 2.

 To reach points in the upper quadrant of the Penrose diagram, one needs a geodesic that begins with a general slope in the $z - t$ plane at its origin point on the right boundary. This is easily obtained, since the action of the special conformal generator, $K^0$ that has $x_*$ as a fixed point changes this slope. Therefore, one may simply consider the operator $e^{i a K^0} \Phi_\textrm{GD}(x_*, z_*) e^{-i a K^0}$.

The crucial fact is that the geodesic gauge fixed kets only span the Hilbert space of states in which the marked geodesic lies along a spatial slice containing the entire $t=0$ slice of the boundary. In the two sided asymptotically AdS context, that includes the left boundary opposite to the right one where the point $x_*$ lies. Since the bulk operator is defined in terms of these states, it includes the projection operator onto the subspace which they span. Such an operator is explicitly left time dependent, so it will fail to exactly commute with $H_L$.

Consider the commutator of left and right boundary framed geodesic observables. This would seem to vanish in all factorized states, since the operators would only act on the left and right Hilbert spaces respectively. However, the appropriately gauged fixed operator acts on both boundaries. One can compute $\<0,0| [\Phi_\textrm{GD, L}(0, z_*), \Phi_\textrm{GD, R}(x, z_*')] |\textrm{tfd}\>$ by finding the non-analytic part of the euclidean path integral with a pair of geodesic insertions, as a function of $x$. This corresponds to the insertion of the geodesics in the geometry shown in Figure 2, and it is clear that they may cross in the interior for sufficiently large $z_*, z_*'$. Therefore the commutator of the gauge fixed operators is nonzero even when acting on the factorized vacuum. It is exponentially small, since it is suppressed by the exponential of the action of the saddle shown in Figure 2. Note that the slice condition for the geodesics is obeyed in the euclidean saddle in this situation, so that the action of the gauge fixed operators is indeed given by a straightforward insertion here.

Furthermore, the action of $\Phi_\textrm{GD}$ on the time shifted states, $|\psi_T \>$, is very different than the naive ungauge fixed observable. The time shifted states are described by the same geometry, but with a shifted origin of time on the left boundary. Therefore for sufficiently large $T$, the left $t=0$ slice will no longer lie on a spatial slice with entirety of the right framed marked geodesic. Thus the operator $\Phi_\textrm{GD}$ as defined above will have a completely different action on those states. In the classical limit, it would simply annihilate them.

This resolves the paradox. To consistently define a bulk effective theory operator, one must refine the standard description to make the operator obey the Hamiltonian constraints nonperturbatively. This involves making various choices; the above is one example. The gauge fixed operators are perfectly well-defined, and no paradox arises for them. On the other hand, their action on most of the exponentially time shifted states differs significantly from the naive description.

\section{Discussion}

In this work, I have argued that bulk observables that have been shown to lack reconstructions as linear CFT operators over the entire range of the bulk effective theory also fail to be nonperturbatively diffeomorphism invariant in the bulk description. This resolves the discrepancy with the gauge/gravity duality, and implies that there is no obstruction to a linear map from the Hilbert space of the bulk effective theory over its entire expected range of validity to the boundary conformal field theory Hilbert space.

A simple example is the connectedness of space in the two sided asymptotically AdS system. It is a well-defined observable in classical general relativity. However, using the description of connected spacetimes as entangled combinations of microstates of the dual CFT, one can show that there cannot exist a linear operator $C$ that measures the connectedness \cite{Bao:2015nca,Berenstein:2016pcx,Berenstein:2017abm}. The solution of the puzzle is that the connectedness of space does not satisfy the Hamiltonian constraints nonperturbatively, and so does not exist as an operator even in the bulk gravity theory.

The bulk physics is often described in terms of boundary relational operators, like a field value at a location determined by some proper distance along a geodesic extending transversally from a given point on the boundary. Such geodesically specified observables are well-defined in classical gravity. Furthermore, in perturbation theory around any given configuration, there is a well-defined quantum operator, which can be represented by a boundary CFT operator in the associated code subspace of states that are perturbative excitations around that fixed state. For example, such bulk observables can be constructed in terms of CFT operators in perturbation theory around exponentially time shifted versions of the eternal AdS black hole in the two sided AdS system \cite{Papadodimas:2015jra,Papadodimas:2015xma}. These states are all described by the same spacetime manifold, with a shift of the origin of time on the left boundary, and thus are in the domain of validity of the long distance bulk effective theory.

However, from the microscopic description of the eternal black hole state as the thermofield double entangled sum of factorized CFT microstates $|E_L\> |E_R\>$, one can show that no linear CFT operator can agree with the expected bulk matrix elements in all of the code subspaces \cite{Marolf:2012xe}.

The resolution is similarly that the usual spacetime path integral formulation of the geodesic defined operators is ambiguous. Moreover, if one defined an operator using bulk perturbation theory around different configurations whose wavefunction is peaked on a ket $|h\>$, then the total object would fail to satisfy the Hamiltonian constraints non-perturbatively.

There exist better ways to define such an operator in the bulk theory, that involve a partially gauge fixed set of kets. These make it well-defined and gauge invariant, but then it does not agree with the results of bulk perturbation theory around all states of the bulk effective theory. In particular, in the operators defined in section 4, there is a projection on to states in which the geodesically specified bulk point is on a spatial slice with the opposite boundary. This results in an order 1 disagreement with the matrix elements of the naive observable in perturbation theory around most time shifted states. In this way, after making such gauge choices to properly define the bulk operator, there is no obstruction to its action as a linear operator in the CFT.

It would be extremely interesting to explore the role of similar constraints in the black hole information paradox, for black holes given by a pure state in global AdS. It is in principle straight forward to express any perturbative bulk calculation in the language of Hartle-Hawking wavefunctionals, as relations acting on the collection of kets on which the configuration is peaked. Presumably, repeating the Hawking argument in this framework simply will not result in a contradiction. But that in itself gives little information, since the expansion of the bulk kets in terms of states corresponding to boundary CFT local operators is very complicated, and cannot be computed in the bulk effective theory.

One surprising fact is that the entire black hole formation and collapse process in AdS can be captured by spatial slices from a fixed time, $t_0$, on the boundary, by taking any boundary spatial slice which is spacelike to the end of the horizon where the black hole fully evaporates. This description of the Hawking process is complimentary to the more standard picture in which one takes gauge fixed spatial slices whose time evolution is tied to the boundary time. Then the process looks like the formation and evaporation of a locally thermalized configuration in the dual CFT, such as a region of deconfined phase in a large $N$ gauge theory. In the CFT description, it is clear that the time evolution is unitary. But the bulk description must already encode the evaporation at a fixed time in the boundary (for which unitarity is tautological, since there is no evolution under the global Hamiltonian).

From that bulk perspective, one has pure constraint equation evolution, and in principle there is a euclidean path integral calculation of the overlaps between the initial kets and the final ones. If the black hole is formed by the collapse of low energy density matter, then the initial wavefunctional is peaked around kets that are in the domain of validity of the bulk effective theory. Similarly, the outgoing Hawking quanta are individually of low energy, and that state will also be in the domain of validity of the effective theory.

The amplitude between such kets at early and late slices (ending at the same boundary time $t_0$) will be exponentially small, since they are connected by some nontrivial euclidean saddle. However, there is an exponentially large number of relevant late time kets, describing the outgoing Hawking quanta, and one expects that the early and late sets of kets are linearly dependent; that is the pure constraint equation evolution. Finding the linear relations again involves partly diagonalizing a matrix of overlaps, which will probably exit the domain of validity of the effective theory. So, unsurprisingly, one does not expect to be able to calculate the exact unitary map from the infalling matter state to the final outgoing state of Hawking quanta from the bulk effective theory alone.

An intriguing possibility is that by taking into account more Planckian configurations, associated to ER=EPR wormholes connecting the black hole interior to the outgoing Hawking quanta \cite{Maldacena:2013xja}, one can use the linear dependence between kets of the resulting different topologies to obtain a more detailed understanding.

The discussions in this work about the restrictions of non-perturbative gauge invariance on bulk operators still leaves unresolved the physical question of measurements in the gravitational bulk. It is well known that there are no local diffeomorphism invariant operators. This certainly does not imply that what we actually measure in gravity are nonlocal observables, in the sense that there is a projection onto eigenspaces of nonlocal operators. The Hamiltonian that describes the measurement process is always the integral of a local density, and the physical measurement process cannot, for example, change the ADM energy encoded by gravitational fields at spatial infinity.

The quantum description of a measurement involves coupling the system that one wishes to measure to an apparatus. This interaction can be written schematically  as $H_\textrm{int} = {\cal O}_\textrm{app} {\cal O}_\textrm{sys}$. Decoherence processes in the apparatus result in a density matrix that is very well approximated by one that is diagonal in the eigenbasis of ${\cal O}_\textrm{sys}$. This approximation can be made parametrically good in the limit of a large apparatus, so the result is equivalent to projection on to an eigenstate of ${\cal O}_\textrm{sys}$, with probabilities given by the Born rule.

In gravity, there is no canonical way of separating the interaction Hamiltonian into diffeomorphism invariant system and apparatus operators, since for a local measurement, it must be defined relationally between them. A useful analogy is an interaction of the form $\psi^\dag_\textrm{app} W \psi_\textrm{sys}$ in electrodynamics, where $W$ is a Wilson line connecting the apparatus and system such that the expression is gauge invariant. There is no canonical way to separate out a gauge invariant ${\cal O}_\textrm{sys}$ - one must add a Wilson line to infinity or otherwise pick a gauge. Moreover, even if one makes such a choice, the dynamics is never well approximated by projection on to eigenstates of ${\cal O}_\textrm{sys}$, since that would affect the electromagnetic field far away. In electrodynamics, the resolution is that one doesn't call such interactions measurements - rather it is an exchange of an electron between the system and apparatus.

However in the bulk gravitational theory, all observations are of this form, and we do consider them to be measurements in practice. Therefore it is necessary to examine the physical measurement and decoherence process directly; there is a irreducible obstruction to replacing it with projection on to eigenstates of an operator.

From this perspective, the problem of finding CFT descriptions of bulk observations is a dynamical one. At large $N$ in certain quantum field theories, one should find that extra non-local operators decohere, ie. that in appropriate states which include a measuring apparatus, there is decoherence in the eigenbasis of such an operator. The only coarse-graining that is intrinsically defined by the system is that given by time averaging. Given a short time scale, $\tau$, one can construct the time dependent density matrix, $$\rho(t) = \int_t^{t+\tau} d t' U(t') |\psi \rangle \langle \psi| U(t')^\dag,$$ where $U(t)$ is the time evolution operator. Then an operator is decohered in the state $|\psi\rangle$ if it approximately commutes with $\rho(t)$ for a long time $T \gg \tau$.\footnote{The fact that the QFT Hamiltonian is local explains why the usual local operators can be made to decohere in this sense.} This, {\it a priori}, depends non-linearly on the state.

The novelty in strongly interacting theories with holographic duals is that, at least around certain states, additional non-local operators can be made to decohere. These correspond to approximately local bulk operators.

For this Lorentzian question, it is possible that MERA-like constructions of the states, and the resulting state dependent operators will play an important role. But it is not clear what is the correct abstract framework to discuss such observables.

Alternatively, it is possible that decoherence simply does not occur in these situations, since there is no factorized subsystem that can be associated to the observer non-perturbatively. This suggests the necessity of a new framework of a different type.

\section*{Acknowledgements} I would like to thank Xi Dong, Monica Guica, Daniel Hawlow, Anton Kapustin, Aitor Lewkowycz, Don Marolf, Kyriakos Papadodimas, Eric Perlmutter, Suvrat Raju, Steve Shenker, Herman Verlinde, Aron Wall, Xi Yin, and especially Juan Maldacena for stimulating and helpful discussions. I also thank Noah Jafferis for creating the figures. This work was supported in part by NSFCAREER grant PHY-1352084 and by a Sloan Fellowship.

\end{document}